\documentclass[aps,prl,twocolumn,superscriptaddress,showpacs]{revtex4}

\usepackage{amssymb}
\usepackage{amsmath}
\usepackage{graphicx}
\usepackage{natbib}
\usepackage{nicefrac}
\usepackage{multirow}
\usepackage{epstopdf}
\usepackage{float}

\newcommand{\ra}{\rangle}

\newcommand{\ket}[1] { |  #1 \ra }

\newcommand{\unit}[1]{\mathrm{\ #1}}

\begin{document}

\title{Fast Room-Temperature Phase Gate on a Single Nuclear Spin in Diamond}
\author{S. Sangtawesin}
\affiliation{Department of Physics, Princeton University, Princeton, NJ 08544, USA}
\author{T. O. Brundage}
\affiliation{Department of Physics, Princeton University, Princeton, NJ 08544, USA}
\author{J. R. Petta}
\affiliation{Department of Physics, Princeton University, Princeton, NJ 08544, USA}
% 03.67.Lx Quantum computation
% 76.30.Mi Color Centers EPR
% 76.70.Fz Dynamic Nuclear Polarization

\begin{abstract}
Nuclear spins support long lived quantum coherence due to weak coupling to the environment, but are difficult to rapidly control using nuclear magnetic resonance (NMR) as a result of the small nuclear magnetic moment. We demonstrate a fast $\sim 500\unit{ns}$ nuclear spin phase gate on a $^{14}$N nuclear spin qubit intrinsic to a nitrogen-vacancy (NV) center in diamond. The phase gate is enabled by the hyperfine interaction and off-resonance driving of electron spin transitions. Repeated applications of the phase gate bang-bang decouple the nuclear spin from the environment, locking the spin state for up to $\sim 140\unit{\mu s}$.
\end{abstract}

\maketitle
The NV center in diamond is one of the most promising systems for quantum computation due to its convenient optical spin initialization and readout schemes, which can be performed at room temperature \cite{Jelezko_PhysRevLett.93.130501,Hanson_PhysRevB.74.161203,Balasubramanian_NMat_8_2009}. In addition to effective manipulation of the electronic spin, hyperfine coupling provides a means to detect and control proximal nuclear spins, enabling multiple qubit operations \cite{Childress_Science_314_2006,Dutt_Science_316_2007,Neumann_Science_320_2008,Jiang_Science_326_2009,Taminiau_PhysRevLett.109.137602,Dreau_PhysRevLett.110.060502,Fuchs_NaturePhys_7_2011}. Recent works have demonstrated high-fidelity initialization and readout of nuclear spins in diamond using an NV center as an auxiliary qubit \cite{Neumann_Science_329_2010,Dreau_PhysRevLett.110.060502}. Robust control of nuclear spins can also be achieved at room temperature with coherence times over one second \cite{Smeltzer_PhysRevA.80.050302,VanDerSar_Nature_484_2012,Maurer_Science_336_2012}.

As qubits, nuclear spins are known for their long coherence times and isolation from the external environment \cite{Saeedi_15112013}. NV centers serve as a perfect gateway for accessing nuclear spins in diamond. However, the magnetic moment of a nuclear spin is roughly 1000 times smaller than that of the electron spin, which imposes a fundamental limit on the relevant interaction time with direct AC magnetic fields. Typically, nuclear spin rotations require at least several microseconds to complete \cite{Dutt_Science_316_2007,Smeltzer_PhysRevA.80.050302}.

In this Letter, we demonstrate fast phase gate operations on a single $I$ = 1 $^{14}$N nuclear spin intrinsic to an NV center by utilizing off-resonant Rabi oscillations of the electronic spin, as previously demonstrated on N@C$_{60}$ fullerene ensembles by Morton \textit{et al.} \cite{Morton_NatPhys_192_2006,Morton_PSSB200669118}. By simultaneously driving electronic transitions associated with the two spin projections of the nuclear qubit, phase accumulation between the nuclear spin states can be generated on the timescale of the electron Rabi oscillations.
Through this quantum control approach, we can achieve a $\pi$-phase gate in less than $500\unit{ns}$, a speed far exceeding that of the nuclear Rabi oscillations $\tau_{n} \sim 40\unit{\mu s}$. The pulses can be applied repeatedly, providing rapid phase shifts to bang-bang decouple the qubit from the environment and preserve the qubit state for as long as $\sim 140\unit{\mu s}$ \cite{Morton_NatPhys_192_2006,Morton_PSSB200669118,Viola_PhysRevA.58.2733}.

\begin{figure}
\centering
\includegraphics[width=\columnwidth]{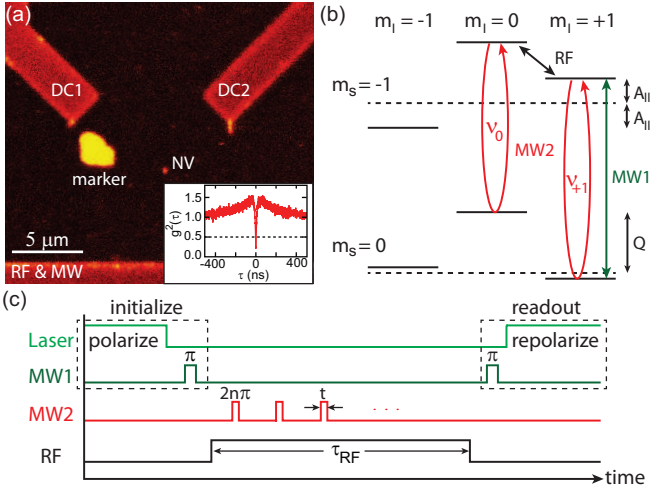}
\caption{(a) Confocal image of the NV center used in the experiment. Two DC electrodes (not used) and a RF \& MW stripline are fabricated near the NV center. Inset: Measurements of the second order correlation function $g^2(\tau)$, with $g^2(0)<0.5$ indicating emission from a single NV center. (b) Energy level diagram, with energy levels indexed by the electron spin quantum number $m_{\rm I}$ and nuclear spin quantum number $m_{\rm S}$. The $m_{\rm S}$ = +1 level is not shown in the level diagram due to the relatively large electron Zeeman energy. Microwave excitations (MW1 and MW2) drive electronic transitions, while RF excitation is used to drive nuclear spin transitions. (c) Pulse sequence used to implement the nuclear spin phase gate during nuclear Rabi oscillations.}
\label{fig:1}
\end{figure}

Our sample is a high purity type IIa diamond (Element Six) with naturally occurring NV centers. We determine the locations of single NV centers relative to pre-patterned alignment marks using fluorescence confocal microscopy, as shown in Fig.\ \ref{fig:1}(a). We verify single photon emission from a single NV center by measuring the photon correlation function $g^2(\tau)$ \cite{Scully_QuantumOptics,som}. After a single NV has been identified, we fabricate on-chip coplanar striplines and DC electrodes near the NV center. The NV used in this experiment has $g^2(0)<0.5$ [see inset of Fig.\ \ref{fig:1}(a)]. Microwave (MW) and radio frequency (RF) pulses are applied to the stripline to drive electron spin and nuclear spin transitions, respectively. The DC electrodes allow Stark shifting of the NV center energy levels \cite{Bassett_PRL_107_2011,Tamarat_PhysRevLett.97.083002}, but are not used in this experiment.

The ground state manifold of the NV center is described by the Hamiltonian:
\begin{align}
H &= D S_z^2 + g_e \mu_B \vec{B} \cdot \vec{S} + \frac{1}{2}A_\perp(S_+I_- + S_-I_+)\notag\\
&\hspace{2cm} + A_\parallel S_z I_z + Q I_z^2 - g_N \mu_N \vec{B} \cdot \vec{I}. \label{eq:nv_hamiltonian}
\end{align}

Here $g_e$ is the electronic g-factor, $\mu_B$ is the Bohr magneton, $\vec{B}$ is the external magnetic field, and $\vec{S}$ ($\vec{I}$) are the electron (nuclear) spin operators. The energy level diagram is shown in Fig.\ \ref{fig:1}(b). Electronic spin levels $m_S=-1,+1$ are separated from $m_S=0$ by a zero-field splitting $D$ = 2.87 GHz. The $m_S=-1$ and $m_S=+1$ energy level degeneracy is lifted by an external magnetic field that results in a Zeeman splitting $g_e\mu_B B_z$, where $g_e \mu_B = 2.802 \unit{MHz/G}$ and the $z$-axis is defined along the NV symmetry axis. The $m_S=+1$ state is far detuned and therefore not shown in the diagram. We use the $m_S = 0$ and $m_S = -1$ levels to encode the electron spin qubit, allowing optical initialization and readout of the NV center electron spin state \cite{Manson_PRB_74_2006,Jelezko_APL_81_2002}. Hyperfine coupling to the $^{14}$N nuclear spin further splits each electronic state into three sublevels, corresponding to the nuclear spin projections $m_I=-1,0,+1$. The $m_I=-1,+1$ states are split from $m_I=0$ by the nuclear quadrupole coupling $Q = -4.962 \unit{MHz}$, and their degeneracy is lifted slightly by nuclear Zeeman splitting, with $g_N \mu_N$ = 0.308 kHz/G. The degeneracy is further lifted by the hyperfine coupling to the electronic spin with $A_\perp = -2.70 \unit{MHz}$ and $A_\parallel = -2.16 \unit{MHz}$ in the $m_S=\pm1$ subspaces \cite{Fuchs_NaturePhys_7_2011,Smeltzer_PhysRevA.80.050302}.

We select the two sublevels $\ket{m_S,m_I} = \ket{-1,+1}$, $\ket{-1,0}$ for our nuclear qubit as the transition is well isolated from the others, allowing selective excitation using the RF field. Moreover, it allows us to perform readout of the nuclear spin state using a microwave $\pi$-pulse (MW1, frequency $\nu_\mathrm{MW1}$), tuned to resonance with $m_I=+1$ transition $\nu_{+1}$, mapping the nuclear spin state to the electronic spin state. The electron spin state is subsequently measured using optical readout \cite{Dutt_Science_316_2007,Smeltzer_PhysRevA.80.050302}. To create well-defined nuclear spin dynamics as a reference, we use RF pulses on resonance with the $^{14}$N transition to induce nuclear Rabi oscillations with Rabi frequency $\Omega_n$. We note that this is necessary as natural Larmor precession of the $^{14}$N nuclear spin is prohibited due to its large quadrupole coupling \cite{Childress_Science_314_2006}.

In detail, when the system is in an arbitrary state $\ket{\psi} = \alpha\ket{-1,0} + \beta\ket{-1,+1}$, with $|\alpha|^2+|\beta|^2=1$, we apply a MW pulse (MW2) with frequency $\nu_\mathrm{MW2}$ for a short duration $t$, driving the electronic transitions $\ket{-1,0}\leftrightarrow\ket{0,0}$ and $\ket{-1,+1}\leftrightarrow\ket{0,+1}$ (with transition frequencies $\nu_0$ and $\nu_{+1}$, respectively). In the strong driving limit, where the electron Rabi frequency greatly exceeds the detunings $\Omega_e \gg \delta_{m_I} \equiv \nu_{\mathrm{MW2}}-\nu_{m_I}$, electrons undergo fast Rabi oscillations between $m_S=0$ and $m_S=-1$ regardless of the nuclear spin state. We ensure that the electron Rabi frequency is the same on both transitions by setting $\nu_\mathrm{MW2}$ midway between $\nu_0$ and $\nu_{+1}$. If $t$ is chosen such that $\Omega_e t = 2n\pi$, the electron spin state will return to the original $m_S=-1$ subspace, with phase accumulation generated by the off-resonant electron spin Rabi oscillations:
\begin{align}
U(t) \ket{\psi} &= e^{in\pi}\left(\alpha e^{i \delta_0 t/2} \ket{-1,0} +\beta e^{i \delta_{+1} t/2} \ket{-1,+1}\right). \label{eq:1}
\end{align}

This process implements a phase gate on the nuclear spin, with the relative phase difference $\Delta\phi = (\delta_{+1}-\delta_0)t/2 = (\nu_0-\nu_{+1})t/2 = A_{||} t/2$ between the two states set by the hyperfine coupling $A_{||}$ and the tunable duration of the MW2 pulse $t$). We note that this process also generates a global phase of $e^{i n\pi}$ on all the states that are being driven by MW2, which can be used as a robust $\pi$-phase gate on the nuclear spins \cite{Waldherr_Nature_2014,Filidou_NatPhys_8_2012}.

The full experimental sequence, including optical initialization and readout, is illustrated in Fig.\ \ref{fig:1}(c). A DC magnetic field $B_z \sim$ 500 G is applied along the NV-axis, bringing the system close to the excited state level anti-crossing (ESLAC) \cite{Fuchs_PhysRevLett.101.117601}. At the ESLAC, the NV center electronic spin is first polarized to $m_S=0$ via optical pumping with a 532 nm laser. During this process, the $^{14}$N nuclear spin is dynamically polarized to $m_I=+1$ \cite{Jacques_PhysRevLett.102.057403,som}. The choice of working at the ESLAC provides high fidelity initialization and readout of the nuclear spin without requiring a Ramsey-type pulse sequence, where the fidelity is limited by a weak selective MW pulse and electron dephasing during the Larmor precession period in the initialization protocol \cite{Dutt_Science_316_2007}. After the system is polarized optically, a selective $\pi$-pulse (MW1) is applied to transfer the population to $\ket{-1,+1}$, completing the initialization process. Spin manipulation is performed by applying a RF pulse with duration $\tau_{\mathrm{RF}}$, while phase gate ``kicks'' from MW2 are simultaneously applied. Finally, optical readout is performed after applying another selective MW1 $\pi$-pulse that converts the population from $\ket{-1,+1}$ to the bright state $\ket{0,+1}$. This yields a photoluminescence (PL) output that is proportional to the $\ket{-1,+1}$ population at the end of the pulse sequence.

\begin{figure}
\centering
\includegraphics[width=\columnwidth]{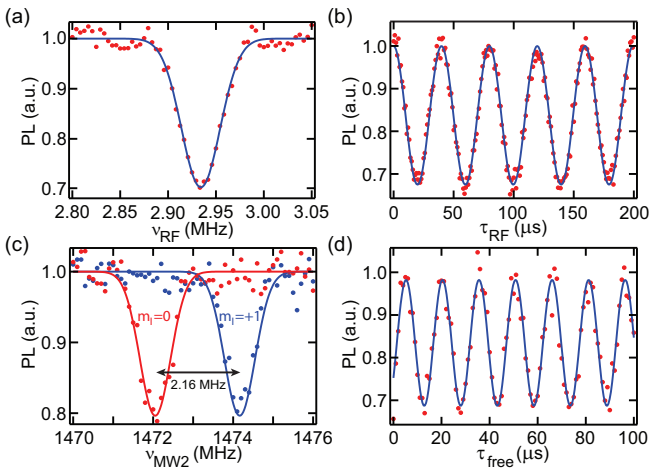}
\caption{(a) PL measured as a function of RF frequency showing a nuclear spin transition at $\nu_\mathrm{RF}$ = $2.934\unit{MHz}$, with a Gaussian fit to the data. (b) PL as a function of RF pulse length $\tau_\mathrm{RF}$ showing nuclear spin Rabi oscillations. (c) PL as a function of selective MW2 $\pi$-pulse frequency $\nu_\mathrm{MW2}$ showing transition frequencies associated with the $m_I = +1$ and $m_I=0$ nuclear spin projections, with Gaussian fits to the data (see main text for details). (d) Nuclear Ramsey measurements: PL measured as a function of free precession time $\tau_\mathrm{free}$ showing no visible decay out to $100\unit{\mu s}$.}
\label{fig:2}
\end{figure}

We probe the $\ket{-1,+1}\leftrightarrow\ket{-1,0}$ nuclear spin transition by sweeping the RF frequency $\nu_{\mathrm{RF}}$ for a fixed $\tau_{\mathrm{RF}}$ that is set to achieve a $\pi$-pulse when on resonance. When $\nu_{\mathrm{RF}}$ is on resonance with the nuclear spin transition, population will be transferred from $\ket{-1,+1}$ to $\ket{-1,0}$. Since $\ket{-1,0}$ is off resonance with MW1, the population transfer leads to a ``dark'' readout and reduces the PL intensity [Fig.\ \ref{fig:2}(a)]. $\nu_{\mathrm{RF}}$ is then tuned to resonance with this transition. By varying $\tau_{\mathrm{RF}}$, nuclear spin oscillations can be observed with a Rabi frequency $\Omega_n \sim 25 \unit{kHz}$ [Fig.\ \ref{fig:2}(b)]. This nuclear Rabi frequency is much greater than the Rabi frequency calculated using only the nuclear gyromagnetic ratio, due to the additional $\sim 10 \times$ enhancement from the electron-nuclear flip-flop, $\Omega_n \approx g_N\mu_N B_{\mathrm{RF}} + A_{\perp} g_e \mu_{B} B_{\mathrm{RF}}/D$ \cite{Childress_Science_314_2006}.

To measure the electronic transition frequency, we perform electron-nuclear double resonance spectroscopy by preparing the system in $\ket{-1,0}$ or $\ket{-1,+1}$ using the calibrated MW1 and RF pulses. A selective MW2 $\pi$-pulse is applied with varying frequency $\nu_\mathrm{MW2}$ and the population remaining in $\ket{-1,0}$ or $\ket{-1,+1}$ is measured by reversing the preparation sequence \cite{som}. A decrease in PL occurs when $\nu_\mathrm{MW2}$ is on resonance with the electron spin transitions [Fig.\ \ref{fig:2}(c)], allowing us to extract the hyperfine coupling $\sim$ 2.16 MHz, consistent with other experiments \cite{Fuchs_NaturePhys_7_2011,Smeltzer_PhysRevA.80.050302,Steiner_PhysRevB.81.035205}. We also implement a Ramsey sequence on the $^{14}$N nuclear spin [Fig.\ \ref{fig:2}(d)]. Here the system is prepared in a superposition state $\ket{\psi}$ = $\frac{1}{\sqrt{2}}(\ket{-1,+1}+\ket{-1,0})$ with a nuclear spin $\frac{\pi}{2}$-pulse and allowed to freely precess for a duration $\tau_\mathrm{free}$ before another $\frac{\pi}{2}$-pulse is applied to rotate the spin back to the original basis \cite{som}. We do not observe any damping in the Ramsey fringes during this measurement interval, indicating a long nuclear spin phase coherence time $T_{2,n}^* > 100\unit{\mu s}$.

\begin{figure}
\centering
\includegraphics[width=\columnwidth]{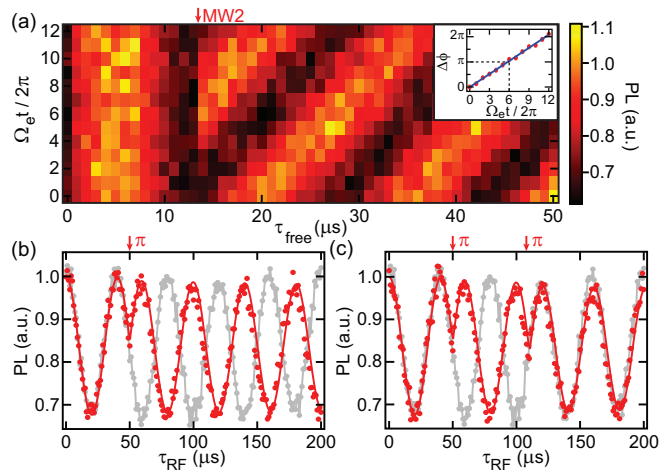}
\caption{(a) Nuclear Ramsey fringes with one phase gate of varying time $t$ applied $13\unit{\mu s}$ into the free-precession interval. Inset: Extracted phase shift of the fringes after the phase gate is applied. (b),(c) Comparison between bare nuclear Rabi oscillations (grey) and nuclear Rabi oscillations with fast $\pi$-phase gates applied (red): (b) One gate applied at $50\unit{\mu s}$. (c) Gates applied at $50\unit{\mu s}$ and $110\unit{\mu s}$. Red curves are simulation results.}
\label{fig:3}
\end{figure}

Nuclear spin phase gates are calibrated by tuning the MW2 frequency $\nu_{\mathrm{MW2}}$ to the midpoint between the $m_I=0$ and $m_I=+1$ transitions. We apply one MW2 pulse of varying duration $t$ during the Ramsey sequence [Fig.\ \ref{fig:3}(a)] and extract the phase of the fringes afterwards. The inset of Fig.\ \ref{fig:3}(a) shows that the nuclear spin phase accumulation $\Delta\phi$ is linearly proportional to the gate duration $t$. For this experiment the MW2 power is tuned such that six full cycles of electron Rabi oscillations correspond to a $\pi$-phase gate ($\Omega_e t/2\pi = 6$). We note that the $\pi$-phase gate duration $t=2\pi/A_{||}=462\unit{ns}$ is fixed by the hyperfine coupling strength $A_{||}=2.16\unit{MHz}$ [Fig.\ \ref{fig:2}(c)].  This corresponds to an electron Rabi frequency $\Omega_e = 12.96\unit{MHz}$.

We now show that the phase gate can be applied during nuclear Rabi oscillations. Results with one and two $\pi$-phase gates are shown in Fig.\ \ref{fig:3}(b) and (c), respectively. The red data points show PL intensity as a function of $\tau_\mathrm{RF}$ with phase gates applied and the red curves are simulation results. For direct comparison, the grey curves show nuclear Rabi oscillations that are not interrupted by phase gates. Phase gates are applied at the times indicated by the red arrows in the figures. The phase shifts $\Delta\phi=\pi$, evident in the two plots, indicate that we successfully applied $\pi$-phase gates on the nuclear spin qubit with each $t = 462\unit{ns}$ gate operation time, far exceeding the speed of nuclear spin Rabi oscillations $\tau_n\sim 40\unit{\mu s}$ shown in Fig.\ \ref{fig:2}(b).

\begin{figure}
\centering
\includegraphics[width=\columnwidth]{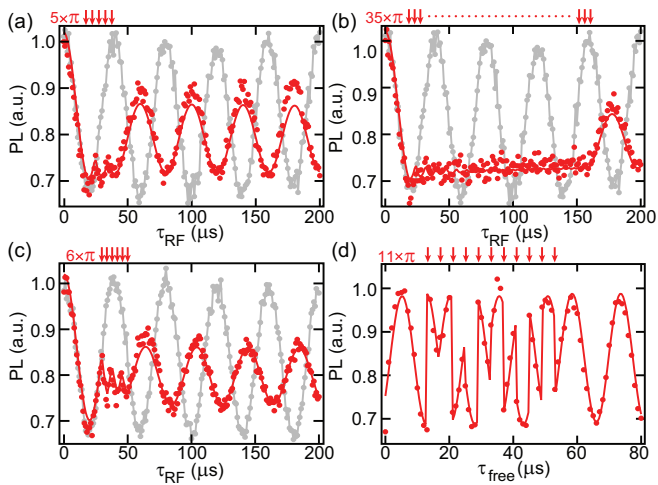}\\
\caption{Comparison between bare nuclear Rabi oscillations (grey) and nuclear Rabi oscillations with multiple $\pi$-phase gates applied in quick succession, locking the nuclear spin (red). Red curves are simulation results. (a) 5 gates applied every $5\unit{\mu s}$ from $20\unit{\mu s}$ to $40\unit{\mu s}$. (b) 35 gates applied every $4\unit{\mu s}$ from $20\unit{\mu s}$ to $156\unit{\mu s}$. (c) 6 gates applied every $4\unit{\mu s}$ from $30\unit{\mu s}$ to $50\unit{\mu s}$. (d) Nuclear Ramsey experiment with 11 $\pi$-phase gates applied at the times indicated by red arrows. In this case, there is no visible decrease in the amplitude of the Ramsey fringes. The theoretical prediction is given by the solid curve \cite{som}.}
\label{fig:4}
\end{figure}

The fast $\pi$-phase gate can be applied repeatedly to decouple the nuclear spin from the RF pulse, effectively locking the nuclear spin state. We demonstrate nuclear spin locking of the $m_I=0$ state by applying multiple $\pi$-phase gates in rapid succession for up to several nuclear Rabi oscillation periods [Fig.\ \ref{fig:4}(a--b)]. We also demonstrate locking of a superposition state of the nuclear spin in Fig.\ \ref{fig:4}(c). While the state preservation is evident, there is $\gtrsim 30\%$ decrease in the amplitude of nuclear Rabi oscillations after several phase gates are applied. The amplitude decrease is asymmetric; it reduces the bright state PL level while leaving the dark state PL level unchanged. This suggests that it is not caused by dephasing of the nuclear spin, in which case the amplitude would be dampened symmetrically. To understand this effect, we first show from simulations that the decrease in contrast is due to the population being driven out of the $\ket{m_S,m_I} = \ket{-1,0},\ket{-1,+1}$ two-level subspace \cite{som}. Then, we argue that this missing population contributes to a ``dark'' readout, resulting in the asymmetric decrease in the oscillations.

To simulate the experiment we use Eq.\ \ref{eq:nv_hamiltonian} and add driving terms:
\begin{align}
 H_N^{\rm ac} &= \Omega_{n} \sin(2\pi\nu_\mathrm{RF} t) I_x \\
 H_e^{\rm ac} &= \Omega_{e} \sin(2\pi\nu_\mathrm{MW2} t) S_x + r \Omega_{n} \sin(2\pi\nu_\mathrm{RF} t) S_z. \label{eq:e-rabi}
\end{align}

The experimentally determined Rabi frequencies ($\Omega_e = 12.96\unit{MHz}$ and $\Omega_n=25\unit{kHz}$) are used in the simulation, with the DC magnetic field fixed at $B_z = 500 \unit{G}$. Here $r$ is a phenomenological parameter to account for an off-axis nuclear drive field that couples to the Zeeman splitting of the electron spin state \cite{som}. Dephasing of the electronic spin is modeled using the Lindblad master equation with the electron spin coherence time $T_2=300\unit{\mu s}$ extracted from spin echo experiments. We start the simulation with the system in a pure state $\ket{0,+1}$ before the MW1 initialization pulse is applied. Electron spin relaxation and nuclear spin dephasing are neglected as there is no observable decay on the timescale of our experiments.

The simulation results indicate that there is a population buildup in the $\ket{m_S,m_I} = \ket{0,0}$ state after the application of multiple phase gates \cite{som}. This population buildup results in a decrease in the maximum population of $\ket{-1,+1}$ Rabi oscillations after the phase gate, which agrees with the experimental result. This is due to a RF drive being applied simultaneously with the MW2 pulses, causing the electronic levels to oscillate at a frequency comparable to the electron Rabi frequency \cite{Childress_PRA82.033839}. A non-ideal rotation on the $\ket{-1,0}\leftrightarrow\ket{0,0}$ transition during the fast phase gates leaves some residual population in $\ket{0,0}$. We account for this effect using the second term in Eq.\ \ref{eq:e-rabi}, which couples the RF drive to the Zeeman splitting of the electron spin state \cite{som}.

Next, we argue that this additional population in $\ket{0,0}$ does indeed contribute to a ``dark'' readout. Typically the $m_S=0$ population would contribute to a ``bright'' readout, as the cycling transition associated with $m_S=0$ does not involve the intersystem crossing through the singlet state $^1A_1$. However, near the ESLAC, $\ket{0, 0}$ is strongly coupled to $\ket{-1, +1}$ in the excited state manifold. In contrast to the experiment performed by Morton \textit{et al.} \cite{Morton_NatPhys_192_2006,Morton_PSSB200669118} where the nuclear spin is initialized using thermal polarization, it is precisely this coupling that makes the nuclear spin polarization possible for our experiment as it allows for Larmor precession between the two states near the ESLAC during optical excitation. Thus, $\ket{0, 0}$ can be converted to $\ket{-1,+1}$ and contribute to a dark state during optical readout due to the non-radiative decay of $\ket{-1,+1}$ through the singlet state. This sequence of events, which is part of the nuclear spin polarization process, occurs on the same $\sim 300\unit{ns}$ timescale as the optical readout \cite{som,Steiner_PhysRevB.81.035205}.

Finally, we show that the population loss is indeed associated with the application of RF pulses. We perform a Ramsey experiment [see Fig.\ \ref{fig:4}(d)] and interrupt the free evolution by applying multiple $\pi$-phase gates. Within our measurement error, there is no visible decay of the Ramsey fringe amplitude after the phase gates are applied. These measurements indicate that there is no population loss after many phase gates have been applied during the time when RF is turned off \cite{som}.

In conclusion, we have demonstrated fast phase gate operations on a nuclear spin qubit in diamond by driving electronic spin transitions of an NV center. $\pi$-phase gates are achieved in 462 ns, approximately 100 times faster than the bare nuclear Rabi frequency. These fast phase gates can be applied repeatedly to preserve the nuclear spin state, providing an alternative method for decoupling the nuclear spin from static environments.

We acknowledge M. V. G. Dutt, G. D. Fuchs, and J. S. Hodges for helpful discussions. Research was supported by the Sloan and Packard Foundations, the National Science Foundation through awards DMR-0819860 and DMR-0846341, and the Army Research Office through PECASE award W911NF-08-1-0189.

\end{document}